\documentstyle[11pt,psfig,a4]{article}
\begin{document}
\title{COLLISIONLESS MODES OF A TRAPPED BOSE GAS}
\author{M.J. Bijlsma and H.T.C. Stoof \\
Institute for Theoretical Physics, University of Utrecht, \\
Princetonplein 5, 3584 CC Utrecht, The Netherlands}
\maketitle
\begin{abstract}
We calculate the excitation frequencies of the 
$m=0$ monopole and $m=2$ quadrupole modes in the
collisionless regime by solving a non-linear 
Schr\"odinger equation for the condensate, 
coupled to a collisionless Boltzmann equation
for the quasiparticles. Since the dynamics of the noncondensate cloud
is also taken into account, the theory satisfies the Kohn theorem.
The spectrum turns out to be strongly temperature dependent
and we compare our results with experiment. 
\end{abstract}

\section{INTRODUCTION}
Since the experimental realization of a Bose condensate
in trapped alkali vapors,
there has been renewed interest into the subject of 
degenerate quantum gases. 
Topics like the equilibrium properties of the condensate,
the dynamics of condensate formation, topological defects, 
and collective excitations have been studied extensively.  
Of particular interest are the collective excitations in the collisionless
regime, because in this regime, where the mean free path of the Bogoliubov 
quasiparticles is much larger than the wavelength of the collective
exitations, there seems to be a 
discrepancy between the experimental 
observations and theoretical calculations.
At temperatures far below the critical temperature 
$T_{c}$, measurements of the low-lying collective excitations 
\cite{jin1,mewes}
are in excellent agreement
with theoretical calculations solving the Gross-Pitaevskii equation 
\cite{stringari,edwards,castin,perez-garcia}, which
describes the condensate dynamics at zero temperature.
At higher temperatures there is a considerable noncondensate fraction, 
and one has to include the mean-field interaction of the thermal cloud
into the evolution equation for the condensate wave function.
Theoretical calculations solving the resulting nonlinear Schr\"odinger 
equation predict almost no temperature dependence of the 
lowest exitation frequencies \cite{dodd,hutchinson1}, 
whereas experiments clearly show a large temperature dependence \cite{jin2}.
This might partly be explained by including in 
the effective interaction between two colliding particles the 
many-body effect of the surrounding gas on the collisions, which
causes the effective two-particle interaction to 
become strongly temperature dependent \cite{bijlsma}.
The frequencies of the low-lying modes will therefore also 
depend on temperature \cite{hutchinson2}. However, in 
these approaches the nonlinear
Schr\"odinger equation describes the dynamics of the condensate in the
presence of a static noncondensed cloud. As a result they violate
the Kohn theorem, which states that there should always be three
center-of-mass modes with the trapping frequencies.
Clearly, this violation is caused by the fact that
we also have to describe the time evolution 
of the thermal cloud. Hence we propose to describe 
the collective excitations in the collisionless regime
by a nonlinear Schr\"odinger equation for 
the condensate wave function that is coupled 
to a collisionless Boltzmann equation describing 
the dynamics of the noncondensed atoms \cite{kirkpatrick,stoof}. 
This resolves our problem, 
because the resulting theory can be shown to
contain the Kohn modes exactly.
A full solution of the collisionless Boltzmann equation 
for the distribution function of the Bogoliubov quasiparticles
is rather complicated. Therefore we apply as a first step in this article 
the Hartree-Fock  approximation for the quasiparticle 
dispersion. This is appropriate 
in the most interesting region, 
near the critical temperature $T_{c}$,
where the mean-field interaction of the
condensate is small compared to the 
average kinetic energy of the noncondensed cloud.
We then determine the eigenfrequencies of the low-lying 
modes by a variational approach
and compare these to the experiments.

\section{COLLISIONLESS DYNAMICS}
Our aim is to describe the coupled dynamics of the condensate
and the thermal cloud in the collisionless regime, where
the wavelength of the collective excitations is much larger
than the mean free path of the quasiparticles.
The time evolution of the condensate 
wave function $\Phi({\bf x},t)$ describes the dynamics of both 
the condensate density $n_{0}({\bf x},t)=|\Phi({\bf x},t)|^{2}$,
and the superfluid velocity 
${\bf v}_{s}({\bf x},t) 
= \hbar \left[ \Phi^{*}({\bf x},t)  \nabla \Phi({\bf x},t)  
- \Phi({\bf x},t)  \nabla \Phi^{*}({\bf x},t) \right] /2im n_{0}({\bf x},t)$.
In the mean-field approximation 
it obeys the nonlinear Schr\"odinger equation, that
includes the effect of the mean-field 
interaction of the noncondensate density $n'({\bf x},t)$, and reads 

\begin{eqnarray}
\label{nlse}
i \hbar \frac{\partial \Phi({\bf x},t)}{\partial t} & = &
\left\{
-\frac{\hbar^{2} \nabla^{2}}{2m} + V^{trap}({\bf x}) - \mu
+ T^{2B} \left[ 2 n'({\bf x},t) + n_{0}({\bf x},t) \right]
\right\} \Phi({\bf x},t) \; .
\end{eqnarray}
Here, $V^{trap}({\bf x})$ denotes the external trapping potential.
The factor of two difference between the condensate and noncondensate
mean-field contibutions results from the  
symmetrisation of the many-body wave function. For the noncondensate part this
contributes a Hartree and a Fock term, but for the condensate part only 
a Hartree term. Moreover, the Hartree and the Fock contributions are equal 
because the two-body interaction is approximated by a hard-core potential 
$V({\bf x}- {\bf x}') = T^{2B} \delta({\bf x}- {\bf x}')$, where
the two-body scattering matrix $T^{2B}=4 \pi \hbar^{2} a/m$ solves
the Lipmann-Schwinger equation for the scattering of two 
particles with zero momentum. In the Thomas-Fermi limit
one can neglect the average kinetic energy 
of the condensate relative to the mean-field interactions,
and the equilibrium density profile of the condensate
is approximately an inverted parabola. In the opposite limit where 
the mean-field interaction is much less than the average
kinetic energy, the result is essentially 
a gaussian density profile associated with the ground 
state of an harmonic oscillator.  

The Boltzmann equation 
is obtained by performing a gradient expansion 
on the equations of motion for the Wigner
distribution, i.e. the Fourier transform of the
one-particle density matrix. This is justified because the 
noncondensate density profile varies on
a much larger length scale 
than the external trapping potential. 
It describes the time evolution
of the quasiparticle distribution function $F({\bf k},{\bf x},t)$,
which gives the noncondensate density profile, 
basically by integrating over momentum space.
The local group velocity
and the local force on a quasiparticle are given by 
the momentum and the spatial derivative of 
the dispersion, respectively. Therefore the Boltzmann equation reads
   
\begin{eqnarray}
\label{cbe}
\left[
\frac{\partial}{\partial t} + 
\frac{\partial \omega({\bf k},{\bf x},t)}{\partial {\bf k}} 
\cdot \frac{\partial}{\partial {\bf x}} -  
\frac{\partial \omega({\bf k},{\bf x},t)}{\partial {\bf x}}  
\cdot \frac{\partial}{\partial {\bf k}}  
\right] 
F({\bf k},{\bf x},t) & = & 
\left[ 
\frac{\partial F({\bf k},{\bf x},t)}{\partial t} 
\right]_{collisions} \; .
\end{eqnarray}
Sufficiently close to the critical temperature, where
$k_{B} T \gg T^{2B} n_{0}({\bf x})$,  
we can treat the quasiparticles in the Hartree-Fock approximation 
and the dispersion is accurately given by

\begin{eqnarray}
\omega({\bf k},{\bf x},t) & = & \frac{\hbar^{2} {\bf k}^{2}}{2m} 
+ V^{trap}({\bf x}) + 
2 T^{2B} \left[ n'({\bf x},t) + n_{0}({\bf x},t) \right] \; .
\end{eqnarray}
The equilibrium distribution can be found by requiring the collision
term to be equal to zero. For the equilibrium density profile the result 
is equivalent to a local density approximation. This is accurate
for the conditions of interest, since also the coherence length of the
gas is much smaller than the the scale on which the equilibrium profile varies.
To find the collisionless exitation spectrum,
we now consider fluctuations around equilibrium. Since 
the eigenfrequencies of the lowest collisionless exitations are much larger
than the average time between two collisions, we can neglect the
contribution from the collisional term. 
The result is thus a linearization of the collionless Boltzmann equation,
which is also known in the context of plasma physics as the Vlasov equation. 

\section{VARIATIONAL APPROACH}
Variational calculations have been quite succesfull in obtaining 
the frequencies of the low-lying modes at zero temperature. Because
of this, and because solving the coupled 
equations (\ref{nlse}) and (\ref{cbe}) is rather difficult, 
we use a variational method to solve these equations.
Therefore we need two variational functions. One for the
condensate wave function, the other for the quasiparticle
distribution. Here, the external trapping potential is 
a quadratic potential 
$V^{trap}({\bf x})= \sum_{i} m \omega_{i}^{2} x_{i}^{2}/2$.
In the experiments performed at JILA and MIT the trap
is cilindrically symmetric, hence
$\omega_{1} = \omega_{2} = \omega_{r}$, and $\omega_{3} = \omega_{z}$. 

First, the condensate wave function is approximated by a gaussian,
\begin{eqnarray}
\Phi({\bf x},t) & = & 
N_{0} \left[ \prod_{i} \left( \lambda_{i} 
\sqrt{\frac{\hbar \pi}{m \omega_{i}}} \right) \right] 
e^{
\sum_{i} \frac{m \omega_{i}}{\hbar}
\left( - \frac{x_{i}^{2}}{2 \lambda_{i}^{2}} +
i \beta_{i} x_{i}^{2} \right)
} \; .
\end{eqnarray}
This is appropriate for the experiments performed 
at JILA which are not in the Thomas-Fermi limit. In fact, the 
ratio of $\hbar \bar \omega$ and $T^{2B} n_{0}$ is about one halve,
where $\bar \omega = (\omega_{r}^{2} \omega_{z})^{1/3}$.
However, choosing the condensate wave function to be a gaussian  
is known to give quantitatively good results
for the condensate modes, even in the Thomas-Fermi limit 
where the ground state is well approximated by an inverted parabola. 
Therefore, we expect the above {\it ansatz} 
to be also at least qualitatively applicable to the MIT experiments.
 
Second, to find an appropriate variational function for the quasiparticle 
distribution function, we take the distribution function 
to be an approximate Maxwell distribution for a gas of non-interacting bosons. 
To describe the monopole and quadrupole modes we 
introduce three scaling paramaters $\{ \alpha_{i} \}$
in the directions $\{ x_{i} \}$ respectively, where $i=\{ 1,2,3 \}$. Hence,
\begin{eqnarray}
F({\bf k},{\bf x},t) & = & 
N' \left[ \prod_{i} \left( \beta \hbar \omega_{i} C_{i} \right) \right]
e^{
- \beta \sum_{i} C_{i} \left[ 
\frac{\hbar^{2} \alpha_{i}^{2}}{2 m} \left(
k_{i} - \frac{m}{\hbar} \frac{\dot \alpha_{i}}{\alpha_{i}} x_{i} 
\right)^{2} + 
\frac{m \omega_{i}^{2}}{2} \frac{x_{i}^{2}}{\alpha_{i}^{2}}
\right]
} \; .
\end{eqnarray}  
This results in a time dependent density profile 
$n'(\{x_{i}/\alpha_{i}(t)\})$. 
For such a profile, the local current should be given by 
$\hbar \langle k_{i} \rangle({\bf x},t) / m = 
\dot \alpha_{i}(t) x_{i} / \alpha_{i}(t)$,
which is correctly reproduced by our trial function. 
Furthermore, 
the equilibrium values for $\alpha_{i}$ are
in general different for $i=\{ 1,2,3 \}$ and larger than  
one, due to the interactions. This 
will cause the equilibruim density profile to broaden, but 
unfortunately also cause the distribution in momentum space
to become anisotropic, which is incorrect because the 
mean-field interactions have no momentum dependence.
The constant overall factors  $\{ C_{i} \}$ are introduced 
to compensate for this
by choosing the value of $C_{i}$ equal to $1/\alpha_{i,e}^{2}$,
where $\alpha_{i,e}$ denotes the equilibrium value of $\alpha_{i}$.

We can now find evolution equations for the variational 
parameters $\{ \beta_{i}, \lambda_{i}, \alpha_{i} \}$
by minimizing the lagrangian for the nonlinear Schr\"odinger
equation and writing the Boltzmann equation as a set of moment equations,
that truncates because of the variational {\it ansatz}. 
Our main result, the equations for 
$\{ \beta_{i} \}$,$\{ \lambda_{i}\}$, and $\{ \alpha_{i} \}$, therefore reads

\begin{eqnarray}
\label{evolution}
\beta_{i} & = & \frac{1}{2 \omega_{i}} 
\frac{\dot \lambda_{i}}{\lambda_{i}} \; , \\ 
\frac{\ddot \lambda_{i}}{\omega_{i}^{2}} + \lambda_{i} & = & 
\frac{1}{\lambda_{i}^{3}} + \sqrt{\frac{2}{\pi}}
\frac{N_{0} a}{l_{i}}
\left(
\frac{l_{i}}{\bar{l}}
\right)^{3}
\left[ 
\prod_{j} \frac{1}{\lambda_{j}} \right]
\frac{1}{\lambda_{i}} + \nonumber \\
&  & 
\frac{4N'}{\pi \sqrt{2 \pi}}
\left(
\frac{\Lambda_{th}}{\bar{l}}
\right)^{5} 
\frac{a}{\bar{l}} \left[ 
\prod_{j} \frac{1}{\beta \hbar \omega_{j} \lambda_{j}^{2} + 
2 \alpha_{j}^{2} \alpha_{j,e}^{2}} \right]
\frac{\lambda_{i}}{\beta \hbar \omega_{i} \lambda_{i}^{2} + 
2 \alpha_{i}^{2} \alpha_{i,e}^{2}} \; ,
\\
\frac{\ddot \alpha_{i}}{\omega_{i}^{2}} + \alpha_{i} & = &
\frac{1}{\alpha_{i}^{3}} + \frac{N'}{8 \pi^{3} \sqrt{2}}
\left(
\frac{\Lambda_{th}}{\bar{l}}
\right)^{5} 
\frac{a}{\bar{l}} \left[ 
\prod_{j} \frac{1}{\alpha_{j} \alpha_{j,e}} \right]
\frac{1}{\alpha_{i} \alpha_{j,e}^{2}} + \nonumber \\
&  & 
\frac{4N_{0}}{\pi^{3} \sqrt{2}}
\left(
\frac{\Lambda_{th}}{\bar{l}}
\right)^{5} 
\frac{a}{\bar{l}} \left[ 
\prod_{j} \frac{1}{\beta \hbar \omega_{j} \lambda_{j}^{2} + 
2 \alpha_{j}^{2} \alpha_{j,e}^{2}} \right]
\frac{\alpha_{i}}{\beta \hbar \omega_{i} \lambda_{i}^{2} + 
2 \alpha_{i}^{2} \alpha_{j,e}^{2}} \; . 
\end{eqnarray}
Here, $\Lambda_{th}$ 
denotes the thermal wavelength $(2 \pi \hbar/m k_{B} T)^{1/2}$,
$\bar{l}$ is equal to $(l_{r}^{2} l_{z})^{1/3}$,
$l_{i}$ denotes
the harmonic oscillator length $(\hbar/m \omega_{i})^{1/2}$, 
and $N_{0}$ and $N'$ denote the total number of 
condensate and noncondensed atoms, respectively.

To determine the spectrum of the low-lying collective excitations
as a function of $T/T_{c}$, we take the critical temperature
and the number of condensate and noncondensed atoms to be given by
the same expressions as those for an ideal gas in a trap, i.e.
$ k_{B} T_{c}=\hbar \bar{\omega}(N/1.202)^{1/3}$,
 $N_{0}= N \left[ 1-(T/T_{c})^{3} \right]$, and  
$N'= N \; (T/T_{c})^{3}$. In addition, we
have included the effects of evaporative cooling, by approximating
the dependence of the total number of particle $N$ on $T/T_{C}$ as found in 
the JILA experiment by a second order polynomial.
We then find the
fixed point of equations (6),(7) and (8) for each temperature and
calculate the eigenvectors and their eigenvalues by linearizing 
around the fixed point. The results are plotted in figure 1. 
 
\begin{figure}[h]
\psfig{figure=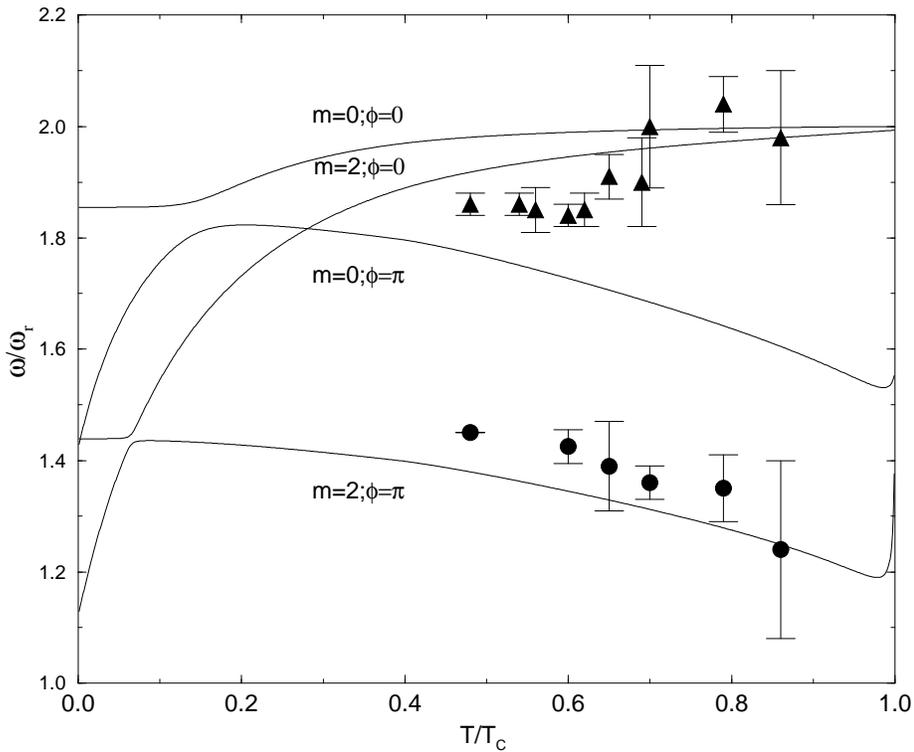}
\label{fig1}
\caption{The low-lying $m=0,2$ collisionless modes as a function of $T/T_{c}$, 
where $\phi$ denotes the relative phase of density profiles of the condensed 
and noncondensed atoms. Also included are the experimental results 
for the m=0 (triangles) and  m=2 (circles) modes 
found in the JILA experiment \cite{jin2}}
\end{figure}

\section{DISCUSSION AND CONCLUSION}
Instead of the usual two modes, found when solving
the Gross-Pitaevskii equation or the nonlinear
Schr\"odinger equation, there
are now four modes. This should not come as a surprise, 
since we are essentially dealing with two gas clouds. 
Without an interaction between these clouds,
there would be a monopole $m=0$ and a 
quadrupole $m=2$ mode for both the condensate and the noncondensed cloud.
If we turn on the interaction these modes get coupled, resulting
in four modes where the condensate and the thermal cloud  
move either in or out of phase. These are the collisionless
analogues of the hydrodynamic first and second sound modes 
\cite{zaremba,kavoulakis1}.
Most importantly, we see from figure 1 that the temperature dependence
of the $m=2$ mode is in good
agreement with the JILA experiment, in contrast with the previous approaches
which do not take into account the dynamics of the noncondensed cloud.
In addition, the $m=0$ mode does have the correct non-interacting limit
near $T_{c}$, but the experimental data drops to the zero-temperature 
limit $(10/3)^{1/2} \; \omega_{r}$ at a higher temperature
than our theoretical curve, which nevertheless shows qualitatively
the same behavior. We therefore believe that the avoided crossing 
between the in and out of phase monopole modes that causes
this behavior, might be the reason for the
strong temperature dependence of the $m=0$ mode found experimentally.
The reason that here the quantitative agreement 
between our theory and the experiments 
is not that good, might be either the use of the Hartree-Fock approximation, 
or of the gaussian approximation for
the condensate wave function. Work to improve on
these approximations is in progress. 
Finally, we note that it might be possible to 
observe also the other two modes experimentally.
Whether this is possible depends
on the overlap of these modes with the 
applied perturbation, and on the damping of the modes, which we have
neglected thusfar. 
However, in principle the collisionless 
Boltzmann equation also contains Landau damping.
Moreover, by including the collision term we should 
be able to describe collisional damping in the same 
way as Kavoulakis, Pethick and Smith \cite{kavoulakis2}.

\section{ACKNOWLEDGMENTS}
\noindent We like to thank C.J. Pethick, E.A. Cornell and F. Langeveld
for usefull discussions.

\end{document}